\documentclass[pra,preprint,superscriptaddress,floatfix,showpacs]{revtex4-1}

\usepackage{amsmath,amsfonts,amssymb}
\usepackage{graphicx}

\def\beq{\begin{equation}}
\def\eeq{\end{equation}}
\def\beqn{\begin{eqnarray}}
\def\eeqn{\end{eqnarray}}

\def\r {{\bf r}}

\begin{document}

\title{Variance as a sensitive probe of correlations\\enduring the infinite particle limit}
\author{Shachar Klaiman}
\email{shachar.klaiman@pci.uni-heidelberg.de}
\affiliation{Theoretische Chemie, Physikalisch--Chemisches Institut, Universit\"at Heidelberg, 
Im Neuenheimer Feld 229, D-69120 Heidelberg, Germany}
\author{Ofir E. Alon}
\email{ofir@research.haifa.ac.il}
\affiliation{Department of Physics, University of Haifa at Oranim, Tivon 36006, Israel}
\date{\today}

\begin{abstract}
Bose-Einstein condensates made of ultracold trapped bosonic atoms have become a central venue in which interacting many-body quantum systems are studied. 
The ground state of a trapped Bose-Einstein condensate has been proven to be 100\% condensed in the limit of infinite particle number and constant interaction parameter [Lieb and Seiringer, Phys. Rev. Lett. {\bf 88}, 170409 (2002)]. The meaning of this result is that properties of the condensate, noticeably its energy and density, converge to those obtained by minimizing the Gross-Pitaevskii energy functional. This naturally raises the question whether correlations are of any importance in this limit. 
Here, we demonstrate both analytically and numerically that even in the infinite particle limit many-body correlations can lead to a substantial modification of the \textit{variance} of any operator compared to that expected from the Gross-Pitaevskii result. 
The strong deviation of the variance stems from its explicit dependence on terms of the reduced two-body density matrix which otherwise do not contribute to the  energy and density in this limit.
This makes the variance a sensitive probe of many-body correlations even when the energy and density of the system have already converged to the  Gross-Pitaevskii result.
We use the center-of-mass position operator to exemplify this persistence of correlations. Implications of this many-body effect are discussed.
\end{abstract}

\pacs{03.75.Hh, 03.65.-w, 05.30.Jp}

\maketitle 

\section{Introduction and Motivation}\label{Intro}

All physical information on the ground state of a quantum system can be obtained from its wavefunction by applying various operators and calculating expectation values. The variance of an operator quantifies to what extent the system under investigation is in an eigenfunction or a superposition of eigenfunctions of the corresponding operator. In this sense it sets the quantum resolution by which the corresponding operator could be measured. 
The variance of single-particle wavefunctions is analyzed in standard quantum mechanics textbooks, see for example \cite{QM_book}. How the \textit{many-body} character of the quantum ground state effects the variance and how this effect persists with increasing the particle number will be the main points addressed in this work. 

Over the past two decades, since they were first experimentally realized \cite{ex1,ex2,ex3}, Bose-Einstein condensates (BECs) made of ultracold trapped bosonic atoms have become a popular ground to study interacting quantum systems, see the reviews \cite{rev1,rev2} and books \cite{book1,book2,book3}. 
There has been tremendous theoretical interest in BECs, and ample studies have been made to describe them using Gross-Pitaevskii, mean-field theory.

Correlations in quantum systems are responsible for many interesting phenomena. 
Generally, one quantifies correlation as the deviation from a reference mean-field picture. 
The broadly accepted paradigm is that, in the limit of large particle number, Gross-Pitaevskii theory properly describes BECs in the ground state. 
To be more precise, the ground state of trapped BECs has been rigorously proven by Lieb and Seiringer to be 100\% condensed in the limit of infinite particle number and constant interaction parameter \cite{Lieb_PRL}. 
Specifically, the energy and density of the condensate were shown to converge to those obtained by minimizing the Gross-Pitaevskii, mean-field energy functional.
This makes the ground state of a BEC a suitable system in which one can study the effects of correlations on different quantum observables and whether such effects persist in the infinite particle number limit. 

In the present work we pose the question whether there are properties
of BECs which, unlike the energy and the density, 
are not obtained in the limit of infinite particle number from the Gross-Pitaevskii, mean-field result.
If there were such properties,
it would imply that there are many-body 
correlations in BECs that are not washed out -- and that can be accessed -- 
in the limit of infinite particle number!
We show both analytically and numerically that the answer to this question is positive and that the variance of any operator is such a property.
The variance can deviate strongly from the Gross-Pitaevskii outcome even in the limit of infinite particle number.
This demonstrates that the variance of an operator
constitutes a sensitive probe for
many-body correlations which are not washed out in this limit.

In Sec.~\ref{Theory} we demonstrate how the energy and the density converge to the Gross-Pitaevskii result while the variance retains the many-body information of the system. 
The results are presented and discussed in Sec.~\ref{Results},
with a sneak preview in Fig.~\ref{f1} below. 
We stress that the conclusion holds true in the infinite particle number limit
which is an exciting, counterintuitive result having a number of immediate and significant consequences.
Of course, these correlations exist for BECs with a finite number of particles 
which are routinely produced in the laboratory.
Summary and outlook are  put forward in Sec.~\ref{Summary}.

\begin{figure}[!]
\includegraphics[width=1.00\columnwidth,angle=0]{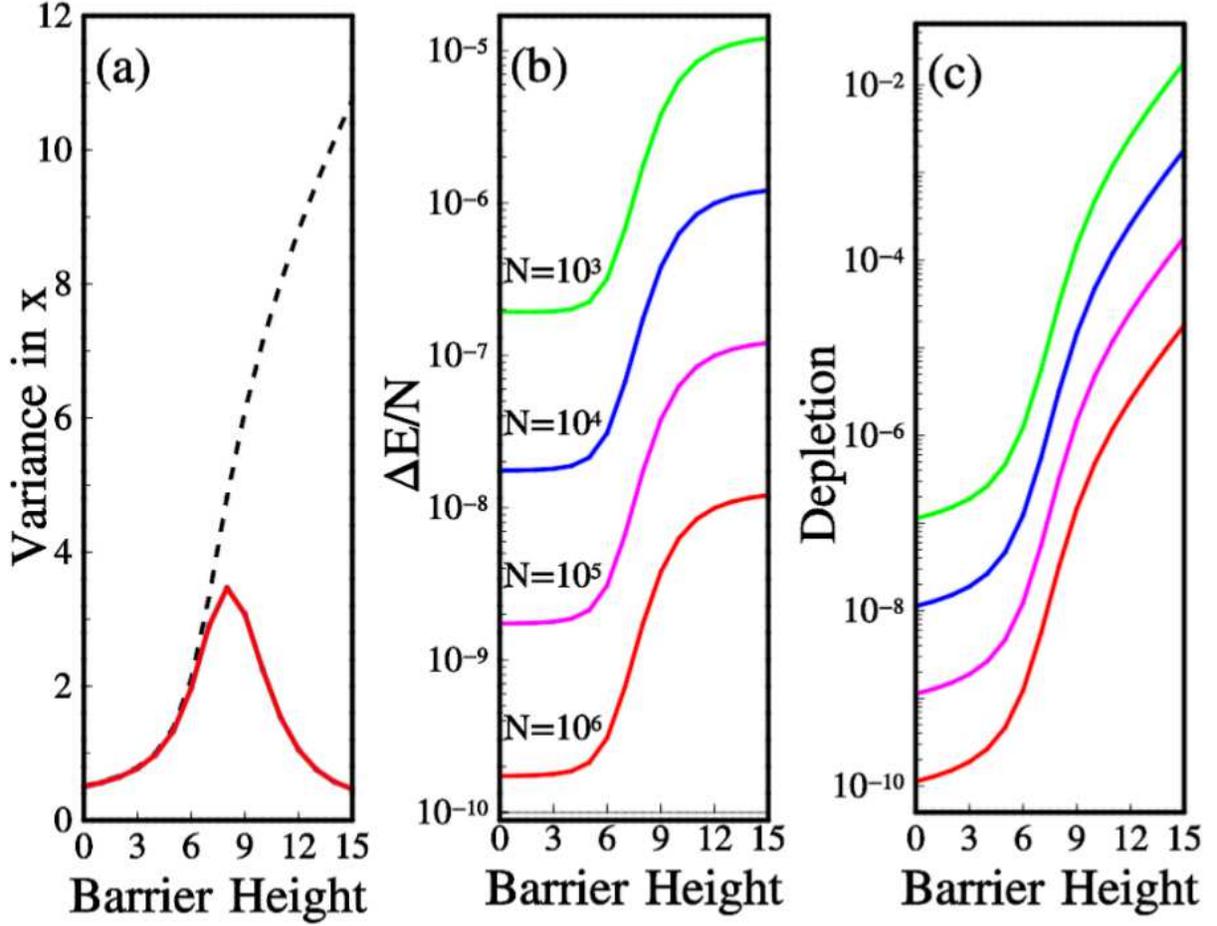}
\caption{(Color online) 
The variance of $\hat X$ of a weakly-interacting Bose-Einstein condensate held in a symmetric trap
for different barrier heights.
Results for $N=1\,000$ (in green), $10\,000$ (in blue), $100\,000$ (in magenta), 
and $1\,000\,000$ (in red) bosons are shown.
The interaction parameter is $\Lambda=\lambda_0(N-1)=0.1$.
(a) Shown is the variance $\frac{1}{N}\Delta_{\hat X}^2$ (full curves)
and the Gross-Pitaevskii variance $\Delta_{\hat x, GP}^2$ in red (in black; dashed curve).
Large differences arise from certain barrier height,
indicating that the correlations term $\Delta_{\hat x, correlations}^2$ becomes dominant.
All four curves for the different numbers of particles
lie atop each other.
(b) The many-body energy and (c) the depletion
are seen to approach and coincide with the Gross-Pitaevskii results,
as is expected from the literature.
Note the small values on the $y$-axes of panels (b) and (c)!
In contrast, the variance converges to a value
substantially different from the Gross-Pitaevskii results for not too shallow barriers.
See the text for further discussion.
The quantities shown are dimensionless.}
\label{f1}
\end{figure}

\section{Theoretical Frameworks}\label{Theory}

\subsection{Trapped Bose-Einstein Condensates in the Infinite Particle Limit}

We begin with the many-body Hamiltonian of $N$ interacting bosons in a trap,
\beq\label{HAM}
\hat H(\r_1,\ldots,\r_N) = \sum_{j=1}^N \hat h(\r_j) + \sum_{j<k} \lambda_0\hat W(\r_j-\r_k).
\eeq
Here $\hat h(\r)$  
is the one-body Hamiltonian,
with kinetic energy and trap potential terms,
and $\hat W(\r_1-\r_2)$ the boson-boson interaction with $\lambda_0$ its strength.

Consider the ground state of the system of bosons,
\beq\label{GROUND}
\hat H(\r_1,\ldots,\r_N) \Psi(\r_1,\ldots,\r_N) = E \Psi(\r_1,\ldots,\r_N),
\eeq
where the wavefunction $\Psi(\r_1,\ldots,\r_N)$ is normalized to one and $E$ is the energy.
We shall find it instructive to continue our discussion
by employing reduced density matrices of $\Psi(\r_1,\ldots,\r_N)$ \cite{Lowdin,Yukalov,RDMs}.

The reduced one-body density matrix is given by
\beq\label{1RDM}
\frac{\rho^{(1)}(\r_1,\r_1')}{N} =
\int d\r_2 \ldots d\r_N \, \Psi^\ast(\r_1',\r_2,\ldots,\r_N) \Psi(\r_1,\r_2,\ldots,\r_N) = 
\sum_j \frac{n_j}{N} \, \alpha_j(\r_1) \alpha^\ast_j(\r'_1).
\eeq
The quantities $\alpha_j(\r)$ are the so-called natural orbitals and $n_j$ their respective occupations
which are used to define the degree of condensation 
in a system of interacting bosons \cite{Penrose_Onsager}.
It is useful, although by no means mandatory, 
to express in what follows quantities using the natural orbitals $\alpha_j$.
The (diagonal of the) reduced two-body density matrix is given by
\beqn\label{2RDM}
\frac{\rho^{(2)}(\r_1,\r_2,\r_1,\r_2)}{N(N-1)} &=& 
 \int d\r_3 \ldots d\r_N \, \Psi^\ast(\r_1,\r_2,\ldots,\r_N) \Psi(\r_1,\r_2,\ldots,\r_N) = \nonumber \\
 &=& \sum_{jpkq} \frac{\rho_{jpkq}}{N(N-1)} \, 
\alpha^\ast_j(\r_1) \alpha^\ast_p(\r_2) \alpha_k(\r_1) \alpha_q(\r_2), 
\eeqn
where  $\rho_{jpkq} = \langle\Psi|\hat b_j^\dag \hat b_p^\dag \hat b_k \hat b_q|\Psi\rangle$
and the creation and annihilation operators are associated with the natural orbitals.
The energy of the ground state is expressed as follows:
\beqn\label{energy_functional}
\frac{E}{N} &=&  \int d\r_1 \, \sum_j 
\Bigg[ \frac{n_j}{N} \cdot \alpha^\ast_j(\r_1) \hat h(\r_1) \alpha_j(\r_1) + \nonumber \\
&+& \sum_{pkq} \frac{\rho_{jpkq}}{N(N-1)} \cdot \frac{\lambda_0(N-1)}{2} \int d\r_2 \, 
\alpha^\ast_j(\r_1) \alpha^\ast_p(\r_2) \hat W(\r_1-\r_2) \alpha_k(\r_1) \alpha_p(\r_2) \Bigg]. \
\eeqn
Note that in Eqs.~(\ref{1RDM}-\ref{energy_functional}) one-body quantities are
divided by $N$ and two-body ones by $N(N-1)$.
The purpose will be seen shortly.
We can now review the properties of the ground state in the infinite particle number limit \cite{Lieb_PRL}
with the help of these reduced density matrices of the system.
 
What is known on properties of the ground state in the infinite particle number limit, i.e., when $N \to \infty$ and  the interaction parameter $\Lambda=\lambda_0(N-1)$ is kept constant?
With respect to the reduced one-body density matrix we have \cite{Lieb_PRL}:
\beq\label{1RDM_LIMIT}
\lim_{N \to \infty} \frac{\rho^{(1)}(\r_1,\r'_1)}{N} = \phi_{GP}(\r_1) \phi^\ast_{GP}(\r'_1).
\eeq
Equation (\ref{1RDM_LIMIT}) means that, in the limit of infinite particle number,
the system of bosons is $100\%$ condensed, i.e.,
$\lim_{N \to \infty} \frac{n_1}{N}=1$.
Side by side, $\lim_{N \to \infty} \alpha_1(\r) = \phi_{GP}(\r)$,
namely, the first natural orbital becomes the Gross-Pitaevskii orbital.
The latter minimizes 
the energy functional,
\beq\label{GP_energy}
\varepsilon_{GP} = \int d\r_1 \, \phi^\ast_{GP}(\r_1) \left[ \hat h(\r_1) + 
\frac{\lambda_0(N-1)}{2} \int d\r_2 \, 
\phi^\ast_{GP}(\r_2) \hat W(\r_1-\r_2) \phi_{GP}(\r_2)\right] \phi_{GP}(\r_1),
\eeq
where $\varepsilon_{GP}$ is the Gross-Pitaevskii energy. 
See for further discussion below. 
The density of the system is the diagonal of the reduced one-body density matrix,
$\rho(\r) = \rho^{(1)}(\r,\r)$.
Then,
\beq\label{DENS_limit}
\lim_{N \to \infty} \frac{\rho(\r)}{N} = |\phi_{GP}(\r)|^2
\eeq
follows straightforwardly from Eq.~(\ref{1RDM_LIMIT}).

With respect to the reduced two-body density matrix we have \cite{Lieb_PRL}:
\beq\label{2RDM_LIMIT}
\lim_{N \to \infty} \frac{\rho^{(2)}(\r_1,\r_2,\r_1,\r_2)}{N(N-1)} = |\phi_{GP}(\r_1)|^2 |\phi_{GP}(\r_2)|^2. 
\eeq
Equation (\ref{2RDM_LIMIT}) means that
$\lim_{N \to \infty} \frac{\rho_{1111}}{N(N-1)} = 1$.

Finally, combining the above infinite particle number limit 
of the reduced one-body and two-body
density matrices, recalling that $\Lambda=\lambda_0(N-1)$ is held fixed,
the energy of the ground state (\ref{energy_functional}) becomes
\beq\label{ENERGY_LIMIT}
\lim_{N \to \infty} \frac{E}{N} = \varepsilon_{GP},
\eeq
where $\varepsilon_{GP}$, the Gross-Pitaevskii energy,
is given in Eq.~(\ref{GP_energy}).

We observe that the analysis of the infinite particle number limit \cite{Lieb_PRL} 
makes use of the {\it dominant} term, $\rho_{1111}$, 
in the expansion of the 
reduced two-body density matrix.
We henceforth pose the question what, if any, would be
the consequence of the other terms?
We would like to stress that such terms 
{\it do not} at all impact the literature result \cite{Lieb_PRL}
on the density, energy, and the $100\%$ condensation of the ground state,
which, in the limit of infinite particle number, are, of course, 
described by the Gross-Pitaevskii results.
But, perhaps, there are other properties
of BECs which are not obtained in the infinite particle number 
limit from the Gross-Pitaevskii result?

\subsection{The \textit{Many-Body} Variance}\label{DaD}

Consider the ground state of $N$ interacting bosons
in a trap described by the wavefunction $\Psi(\r_1,\ldots,\r_N)$.
Let $\hat A(\r)$ be an hermitian operator.
In order to make our presentation more concrete, we choose $\hat A(\r)=\hat x$, 
namely the position operator in the $x$ direction.
Other operators representing different quantities, for instance, the momentum operator,
are informative as well and the derivations below can be easily adapted. 
The choice of the position operator will also allow us for some immediate conclusions.

We begin with the center-of-mass position operator,
\beq\label{lin_term}
 \hat X = \sum_{j=1}^N \hat x_j.
\eeq
The operator $\hat X$ is referred to as a one-body operator 
acting in the space of $N$ particles.

A straightforward calculation gives 
the average of the center-of-mass position operator in the ground state,
$\frac{1}{N}\langle\Psi|\hat X|\Psi\rangle = \int d\r \frac{\rho(\r)}{N} x$,
which is 
seen to be directly related to the density of the system.
Using Eq.~(\ref{DENS_limit}) we can readily obtain that
\beq\label{CM_X_LIMIT}
\lim_{N \to \infty} \frac{1}{N}\langle\Psi|\hat X|\Psi\rangle = \int d\r |\phi_{GP}(\r)|^2 x.
\eeq
Equation (\ref{CM_X_LIMIT}) states 
that, in the infinite particle number limit,
the outcome of an average of a one-body operator in the ground state
is given by the Gross-Pitaevskii result.

But what about the variance of a one-body operator?
To compute the variance we need the square of the center-of-mass position operator which is given by
\beq\label{sqr_term}
 \hat X^2 = \sum_{j=1}^N \hat x_j^2 + \sum_{j<k} 2 \hat x_j \hat x_k
\eeq
and seen to be comprised of one-body {\it and} two-body operators.
The presence of the two-body operator in (\ref{sqr_term}) might suggest that the variance can capture many-body correlations in the interacting system.
Recall, however, that the Hamiltonian (\ref{HAM}) also contains a two-body operator, the boson-boson interaction.
Yet, no many-body correlations are captured in the limit of infinite particle number by the energy [see Eq.~(\ref{ENERGY_LIMIT})], so what could be the difference here?

Expressed in terms of the above quantities, the variance of the center-of-mass position operator of the system takes on the form
\beqn\label{dis}
\frac{1}{N}\Delta_{\hat X}^2 &=& \frac{1}{N} 
\left(\langle\Psi|\hat X^2|\Psi\rangle - \langle\Psi|\hat X|\Psi\rangle^2\right) = \nonumber \\
&=& \int d\r \frac{\rho(\r)}{N}x^2  - N \left[\int d\r \frac{\rho(\r)}{N}x\right]^2 + \nonumber \\
&+& \sum_{jpkq} \frac{\rho_{jpkq}}{N(N-1)} \cdot (N-1) \int d\r_2 \, 
\alpha^\ast_j(\r_1) \alpha^\ast_p(\r_2) \, x_1x_2 \, \alpha_k(\r_1) \alpha_q(\r_2). \
\eeqn
Equation (\ref{dis}) deserves a closer look. 
The variance $\frac{1}{N}\Delta_{\hat X}^2$ consists of two kinds of terms,
one-body terms
and a two-body term.
We will study the impact and interplay of the one-body and two-body terms.
Crucially, the latter is {\it enhanced} by the factor $(N-1)$ which 
will lead to intriguing results in the limit of infinite particle number.
We note that, in the absence of boson-boson interaction, 
the variance $\frac{1}{N}\Delta_{\hat X}^2$ boils down 
to that of a single particle.

We can now turn to the infinite particle number limit of Eq.~(\ref{dis}) which gives the final expression
\beqn\label{dis_THERMO}
& & \lim_{N \to \infty} \frac{1}{N}\Delta_{\hat X}^2 = 
\Delta_{\hat x, GP}^2 + \Delta_{\hat x, correlations}^2, \nonumber \\ 
& & \qquad \Delta_{\hat x, GP}^2 = 
\int d\r |\phi_{GP}(\r)|^2 x^2 - \left[\int d\r |\phi_{GP}(\r)|^2 x\right]^2, \nonumber \\
& & \qquad \Delta_{\hat x, correlations}^2 =
\lim_{N \to \infty} 
\sum_{jpkq \ne 1111}\frac{\rho_{jpkq}}{N} \int d\r_2 \, 
\alpha^\ast_j(\r_1) \alpha^\ast_p(\r_2) \, x_1x_2 \, \alpha_k(\r_1) \alpha_q(\r_2). \
\eeqn
In obtaining Eq.~(\ref{dis_THERMO}) we have made used
of the relations $\lim_{N \to \infty} \alpha_1(\r) = \phi_{GP}(\r)$ 
and $\lim_{N \to \infty} \frac{\rho_{1111}}{N(N-1)} = 1$ discussed above.
We see that, in the infinite particle number limit, the variance can
be expressed as the Gross-Pitaevskii variance, $\Delta_{\hat x, GP}^2$, plus a
second term which we call the correlations term, $\Delta_{\hat x, correlations}^2$.
This is already fundamentally different than the situations for the quantities described so far,
namely, the energy, density, and condensation.
The question that arises is what is the nature of the second term in Eq.~(\ref{dis_THERMO})?
A central matter to pay attention 
to is that
the summation therein 
only includes terms where at least one 
of the indexes is a {\it higher} natural orbital (i.e., second or above).
Hence, $\Delta_{\hat x, correlations}^2$ only contributes in the presence of 
occupation of natural orbitals other than the first natural orbital (the so-called condensed mode).
To remind, the depletion of the BEC is given by the occupation of all but the first natural orbital divided by $N$.
Clearly, in the limit of infinite particle number the depletion becomes infinitesimal. 
Still the question remains whether a tiny occupation of the higher natural orbitals can make a difference?
Yes, it can.
Does this contradict the literature results for the energy, density, and condensation \cite{Lieb_PRL}?
No, it does not. 
Let us see how.

The occupation of the first natural orbital is of order $\mathcal{O}(N)$,
and the (sum of the) rest is of order $\mathcal{O}(1)$.
This is compatible with the literature result 
that $100\%$ condensation and the Gross-Pitaevskii density 
are recovered in the limit of infinite particle number, see Eqs.~(\ref{1RDM_LIMIT}) and (\ref{DENS_limit}).
The first matrix element of the reduced two-body density matrix,
$\rho_{1111}$, is of order $\mathcal{O}(N^2)$,
and the (sum of the) rest is of order $\mathcal{O}(N)$.
This is compatible with the literature result 
that the Gross-Pitaevskii energy is recovered in the infinite particle number limit, see Eq.~(\ref{ENERGY_LIMIT}).
But, unlike the case of the energy which is scaled by the interaction parameter $\Lambda=\lambda_0(N-1)$,
the variance is {\it enhanced} by the factor $(N-1)$.
This allows the variance to pick-up, in the limit of infinite particle number, 
information from the other terms of the reduced two-body density matrix
that do not contribute to the above quantities!

We begin with an analytical case.
Consider interacting bosons in an harmonic trap, an isotropic one without loss of generality. 
The one-body Hamiltonian is $\hat h(\r) = -\frac{1}{2}\big(\frac{\partial^2}{\partial x^2}+\frac{\partial^2}{\partial y^2}+\frac{\partial^2}{\partial z^2}\big) + \frac{1}{2} \omega^2 \r^2$,
where $\omega$ is the frequency of the trap and $\hbar=m=1$.
The details of the boson-boson interaction, namely its strength and shape, 
are not important for our discussion.
It is well known that one can separate the center-of-mass coordinate, $\frac{1}{\sqrt{N}}\sum_{j=1}^N \r_j$,
from the remaining, $(N-1)$ relative-motion coordinates.
Consequently, the $N$-boson wavefunction becomes a product of
the center-of-mass wavefunction,
$\left(\frac{\omega}{\pi}\right)^{3/4} e^{-\frac{\omega}{2}(\frac{1}{\sqrt{N}}\sum_{j=1}^N \r_j)^2}$,
and a relative-motion wavefunction of the respective $(N-1)$ coordinates.
It goes without saying that, due to this separability, the variance takes on the simplest form,
\beq\label{VAR_HO}
\frac{1}{N}\Delta_{\hat X}^2 = \frac{1}{2\omega}, \qquad \forall N,
\eeq 
i.e., irrespective to the number of particles $N$.
Eq.~(\ref{VAR_HO}) thus holds (also) 
in the infinite particle number limit.

On the other hand, it is well known that a repulsive boson-boson interaction
broadens the Gross-Pitaevskii orbital, $\phi_{GP}(\r)$, in comparison with
that of the non-interacting system \cite{rev1,rev2,book1,book2,book3}.
Generally, the Gross-Pitaevskii orbital depends on the interaction parameter, $\Lambda=\lambda_0(N-1)$, and on its shape.
Thus, we arrive at the observation that 
$\Delta_{\hat x, GP}^2$ for repulsive bosons in an harmonic trap
is always larger than $\frac{1}{N}\Delta_{\hat X}^2$,
unless the interaction is zero, and then the 
occupation of the higher natural orbitals is strictly zero.
As a consequence of this analysis,
we see that it is the second term in Eq.~(\ref{dis_THERMO}), 
which is non-vanishing and negative,
that accounts for
the {\it always existing} 
many-body correlations in the system 
leading to the value (\ref{VAR_HO}) of the variance.
To remind, the variance (\ref{VAR_HO}) is independent
of the interaction strength
and equals that of a single particle in an harmonic potential \cite{QM_book}.
Thus,
the relation 
\beq\label{CORR_HAM}
\Delta_{\hat x, correlations}^2 = \frac{1}{2\omega} - \Delta_{\hat x, GP}^2,
\eeq
which is easily computable for a general system of interacting bosons in an harmonic potential,
can be used to quantify these correlations in the limit of infinite particle number.

\section{Numerical Investigations and Discussion}\label{Results}

The results of the previous Subsec.~\ref{DaD},
when put in the context of the infinite particle number limit,
are somewhat surprising.
We have seen in the analytical example why the variance {\it is not}
given in the limit of infinite particle number by the Gross-Pitaevskii result.
What happens to the variance when the bosons are trapped
in an unharmonic potential, where the center-of-mass is not separable?
In this case an analytical treatment is generally excluded
and one is opt to resort to numerical investigations.

\begin{figure}[!]
\includegraphics[width=0.90\columnwidth,angle=0]{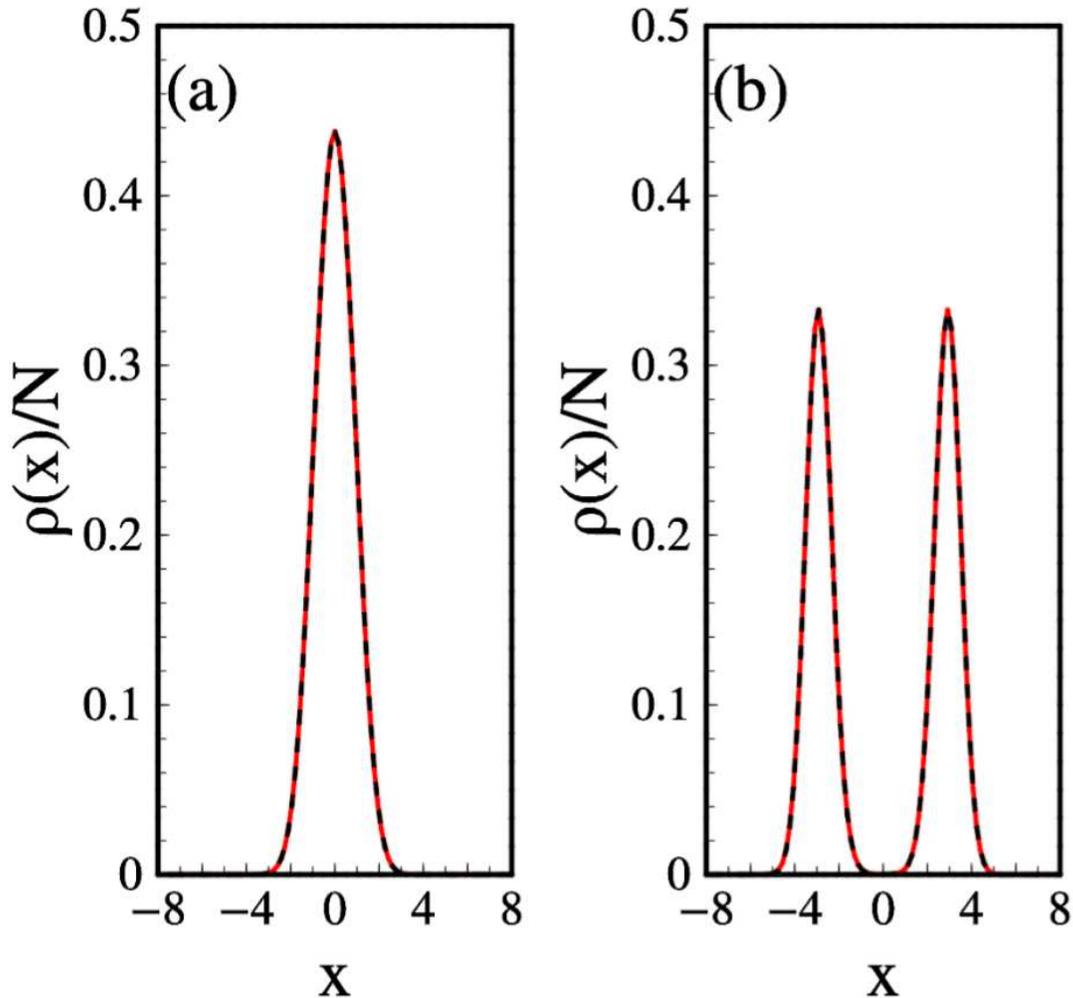}
\caption{(Color online) 
The density of a Bose-Einstein condensate held in a symmetric trap.
The barrier height is (a) A=3 and (b) A=12.
The number of bosons is $N=100\,000$ and the interaction parameter is $\Lambda=\lambda_0(N-1)=0.1$.
The density (in red; full curve) coincide with the Gross-Pitaevskii limit (in black; dashed curve),
as is expected from the literature,
whereas the respective variances deviate substantially from each other,
as seen in Fig.~\ref{f1}.
See the text for more details.
The quantities shown are dimensionless.
}
\label{f2}
\end{figure}

We would now like to investigate numerically
the variance of the center-of-mass position operator $\hat X$ 
of a BEC held in an unharmonic trap potential.
We need a suitable and accurate many-body tool
to arrive at detailed conclusions.
Such a many-body tool is the multiconfigurational time-dependent
Hartree for bosons (MCTDHB) method,
that has been well documented \cite{MCTDHB1,MCTDHB2,book_MCTDH,book_nick},
extensively applied \cite{BJJ,MCTDHB_OCT,MCTDHB_Shapiro,LC_NJP,Breaking},
and benchmarked \cite{Benchmarks} in the literature.
To obtain the many-body ground state we propagate the MCTDHB equations-of-motion in imaginary time
\cite{Benchmarks,MCHB,MCTDHB_3D_stat}.
For the computation details, please see the appendix.

We begin with a symmetric system in one spatial dimension.
Specifically, we consider $N=1\,000$, $10\,000$, $100\,000$, and $1\,000\,000$ 
bosons in a double-well potential.
The one-body Hamiltonian, taken from Ref.~\cite{RDMs},
is $\hat h(x) = -\frac{1}{2}\frac{\partial^2}{\partial x^2} + \frac{x^2}{2} + Ae^{-x^2/8}$.
For $A=0$ there is no barrier,
the trap potential is harmonic,
and the variance is $\frac{1}{2}$ for any number $N$ of particles, 
see Eq.~(\ref{VAR_HO}). 
We would like to study the effect of the barrier on the variance
for up to one million trapped bosons,
and see what can be concluded on the variance in the 
infinite particle number limit.
The interaction is taken to be the standard contact interaction, 
$\hat W(x_1-x_2)=\delta(x_1-x_2)$,
with the interaction parameter
$\Lambda=\lambda_0(N-1)=0.1$ held fixed.
This guarantees that for any number of particles $N$ the Gross-Pitaevskii, 
mean-field solution is the same.
The results are collected in Fig.~\ref{f1}.

Fig.~\ref{f1}a shows the variance $\frac{1}{N}\Delta_{\hat X}^2$
and the Gross-Pitaevskii variance $\Delta_{\hat x, GP}^2$ for 
$N=1\,000$, $10\,000$, $100\,000$, and $1\,000\,000$ bosons.
We can see that for low barrier heights
the two quantities essentially coincide,
whereas substantial differences emerge for higher barrier heights.
As the number of particles increases, 
the difference
between the ground-state energy and the Gross-Pitaevskii energy
decreases, see Fig.~\ref{f1}b.
The depletion of the BEC decreases similarly, see Fig.~\ref{f1}c.
Note the small values on the $y$-axes of Figs.~\ref{f1}b and \ref{f1}c!
Side by side, the density of the BEC and the Gross-Pitaevskii density
coincide, see Fig.~\ref{f2}.
The behavior of the energy, the depletion, and the density 
seen in Figs.~\ref{f1} and \ref{f2} are in line with the literature 
result on the infinite particle number limit.
We can see that with increasing of the barrier
height the depletion increases,
and so does the deviation of the variance $\frac{1}{N}\Delta_{\hat X}^2$
from the Gross-Pitaevskii variance $\Delta_{\hat x, GP}^2$.
This manifests the role of even a tiny amount of population outside the condensed mode
leading to a strong impact on the variance, see Eq.~(\ref{dis_THERMO}).

Returning to Fig.~\ref{f1}a,
in contrast to the behavior of the depletion, the energy, and the density,
the curves of the variance for different $N$ are seen to lie atop each other.
The latter indicates that the correlations term $\Delta_{\hat x, correlations}^2$
dominates the variance in the infinite particle number limit.
An interesting point to take from Figs.~\ref{f1}b and \ref{f1}c is
how the limit of infinite particle number is approached with increasing $N$.
The respective curves are seen to be shifted vertically from each other.
This indicates that, 
whereas the depletion decreases in the infinite particle number limit, 
the total number of particles residing in the higher 
natural orbitals is constant.
When enhanced by the number of particles $(N-1$) in the reduced two-body density matrix,
this leads to the macroscopic difference of the variance from the Gross-Pitaevskii result, see Fig.~\ref{f1}a.

To shed more light on the physical meaning of the variance $\frac{1}{N}\Delta_{\hat X}^2$ 
we examine the densities shown in Fig.~\ref{f2} more closely.
For barrier height $A=3$ the trap is essentially a single well.
Accordingly, there is a single density peek, see Fig.~\ref{f2}a.
The depletion for $N=100\,000$ bosons is just above $10^{-9}$
and consequently, the correlations part $\Delta_{\hat x, correlations}^2$ is very small.
In this case, the variance essentially equals the Gross-Pitaevskii result,
and both match the full-width-half-maximum of the density
($\sim 2\sqrt{\frac{1}{N}\Delta_{\hat X}^2}$).
When the barrier height is $A=12$,
the density is separated into two parts, see Fig.~\ref{f2}b,
yet the system is essentially condensed,
with depletion of slightly above $10^{-5}$.
This amounts to about $1$ atom outside the condensed mode
(the first natural orbital)!
But that is enough,
due to the {\it enhancement} by $(N-1)$ in the correlations term $\Delta_{\hat x, correlations}^2$,
the variance now differs substantially from the Gross-Pitaevskii limit.
Interestingly, the variance
matches the the full-width-half-maximum of each density peek
($\sim 2\sqrt{\frac{1}{N}\Delta_{\hat X}^2}$),
whereas the Gross-Pitaevskii limit
corresponds to the distance between the 
density peeks ($\sim 2\sqrt{\Delta_{\hat x, GP}^2}$).
This is quite an effect of a single excited atom.
And it persists in the limit of infinite particle number,
because there will always be that excited atom,
enhanced by the $(N-1)$ factor!

\begin{figure}[!]
\includegraphics[width=1.00\columnwidth,angle=0]{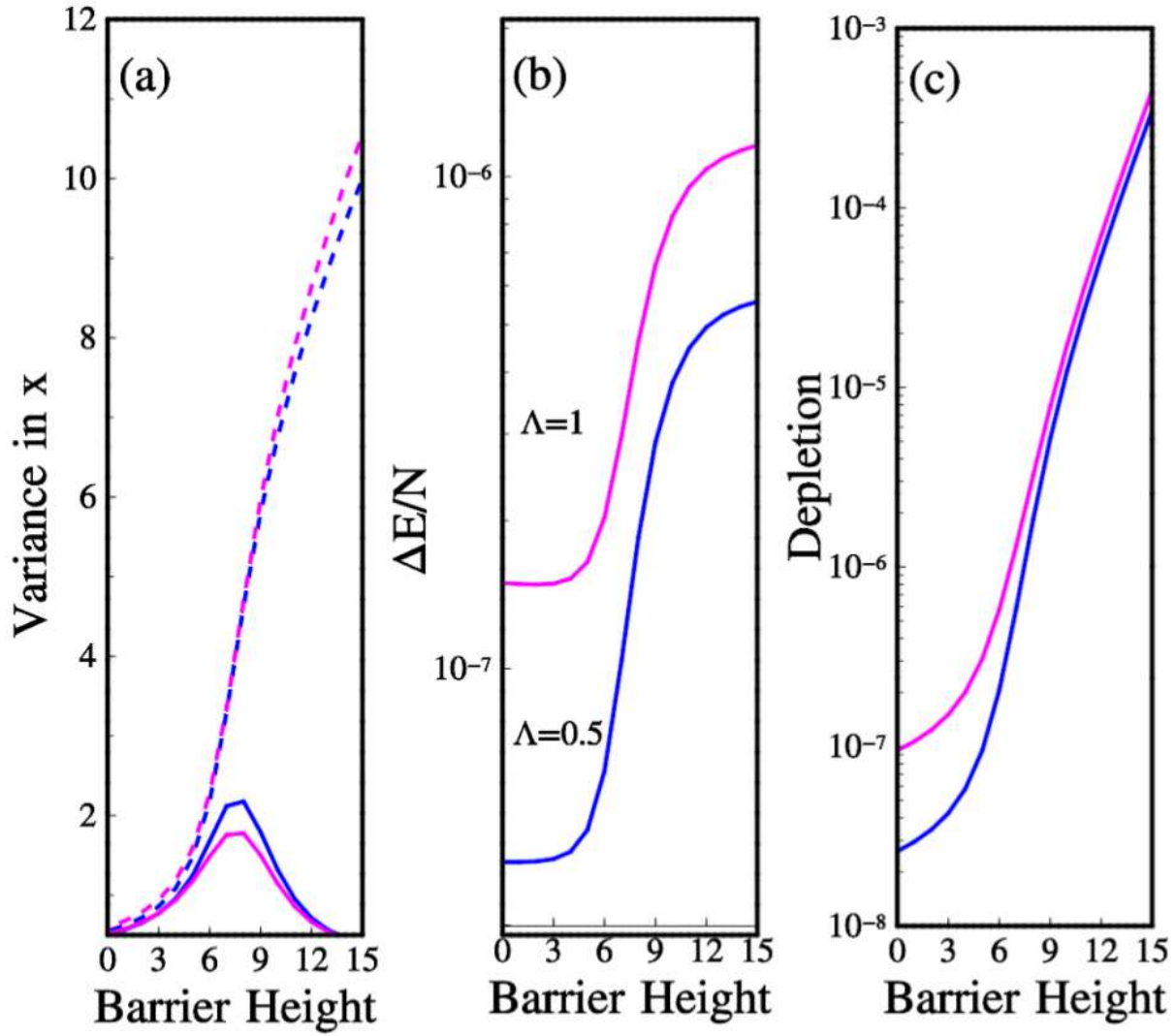}
\caption{(Color online)
Same as Fig.~\ref{f1} but for an asymmetric trap.
Results for $N=100\,000$ and the two interaction parameters
$\Lambda=\lambda_0(N-1)=0.5$ and $1$ are shown.
Note the small amount of depletion.
The variance behave qualitatively the same as in the symmetric trap,
compare to Fig.~\ref{f1},
and deviates substantially from the Gross-Pitaevskii limit for not
too shallow barriers.
The quantities shown are dimensionless.}
\label{f3}
\end{figure}

Finally, the results plotted in Figs.~\ref{f1} and \ref{f2} are 
obtained at the level of $M=2$ orbitals in the MCTDHB method
and found to be converged.
For details of the convergence of the variance with 
increasing the number of orbitals $M$ please see the appendix. 

\begin{figure}[!]
\includegraphics[width=0.90\columnwidth,angle=0]{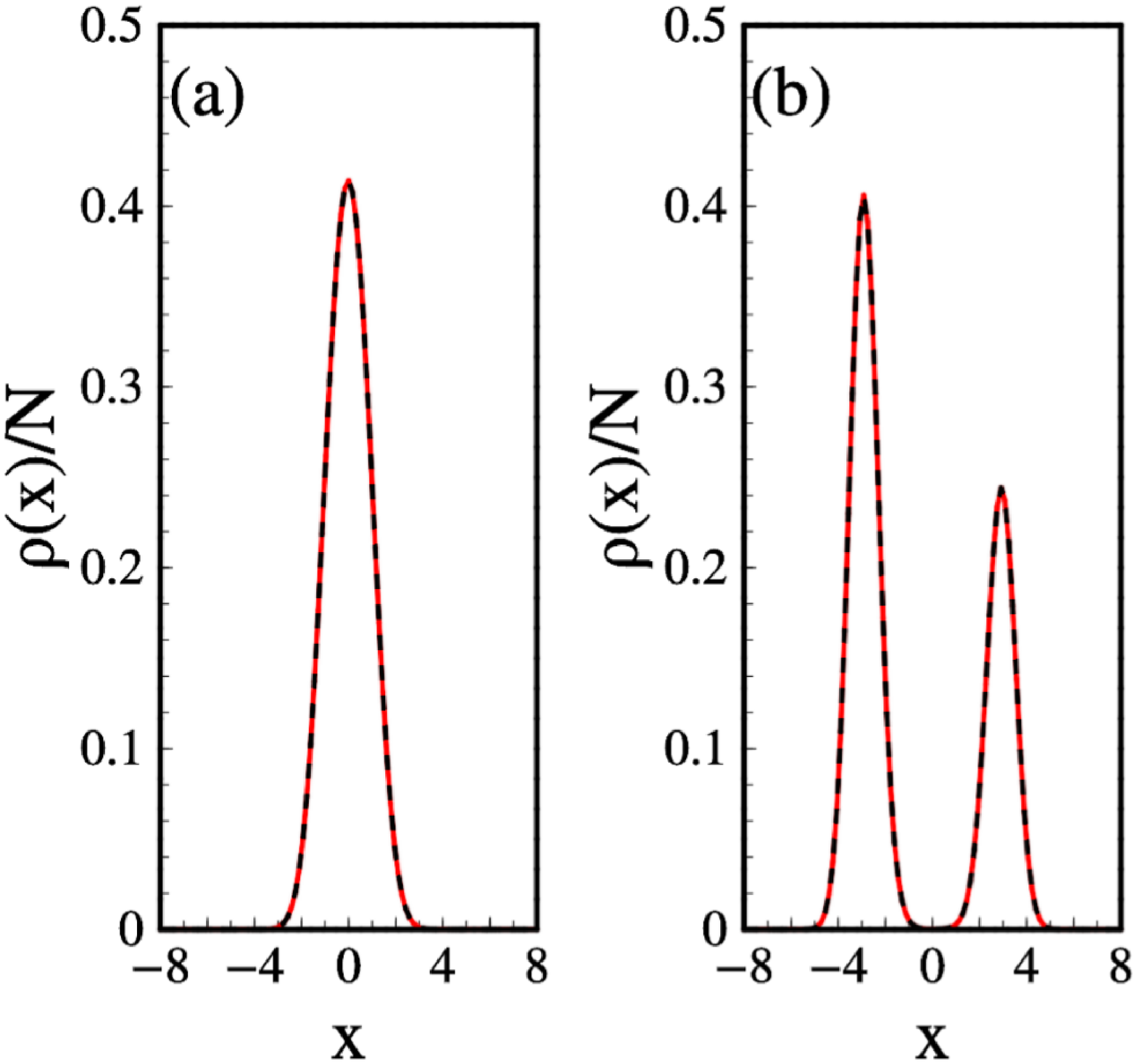}
\caption{(Color online) 
Same as Fig.~\ref{f2} but for an asymmetric trap.
The number of bosons is $N=100\,000$ 
and the interaction parameter is $\Lambda=\lambda_0(N-1)=0.5$.
The density and the Gross-Pitaevskii result coincide.
Compare to the respective variances in Fig.~\ref{f3}.
The quantities shown are dimensionless.
}
\label{f4}
\end{figure}

We have so far treated the variance of $\hat X$ in a symmetric trap;
Whether the harmonic trap, which admits an analytical treatment,
or a double-well trap, which requires a numerical investigation.
When there is spatial symmetry in the system the natural orbitals
possess symmetry as well, e.g., they are even and odd functions
in case of a one-dimensional symmetric trap.
Consequently, there are terms of the reduced
two-body density matrix that vanish due to the spatial symmetry of the system.

As we have seen above, the difference in the variance in the infinite particle number limit from the Gross-Pitaevskii result has to do
with excitations of atoms to higher natural orbitals.
No argumentation based on spatial (or other) symmetry was used, hence we do not expect this many-body effect to depend on symmetry. Still, the interplay of the different terms in Eq.~(\ref{dis_THERMO}) varies in the asymmetric case which warrants an example. 

The one-body Hamiltonian now takes the from $\hat h(x) = -\frac{1}{2}\frac{\partial^2}{\partial x^2} + \frac{x^2}{2} + Ae^{-(x-0.01)^2/8}$.
The interaction is taken again to be $\hat W(x_1-x_2)=\delta(x_1-x_2)$ with the interaction parameters $\Lambda=0.5$ and $1$.
The number of bosons is $N=100\,000$.
The results are collected in Figs.~\ref{f3} and \ref{f4} 
and show that lifting the spatial symmetry does not change the conclusion that the variance of $\hat X$ of a trapped BEC
can differ substantially from the Gross-Pitaevskii limit.
Note that, despite the apparent slight asymmetry of the potential, the density is quite imbalanced between the two wells, see Fig.~\ref{f4}b.

As a final example, we consider a finite BEC made of $N=100$ bosons in a three-dimensional, unharmonic trap.
We would like to touch upon how unharmonicity and interaction conspire to vary the variance.

\begin{figure}[!]
\includegraphics[width=1.00\columnwidth,angle=0]{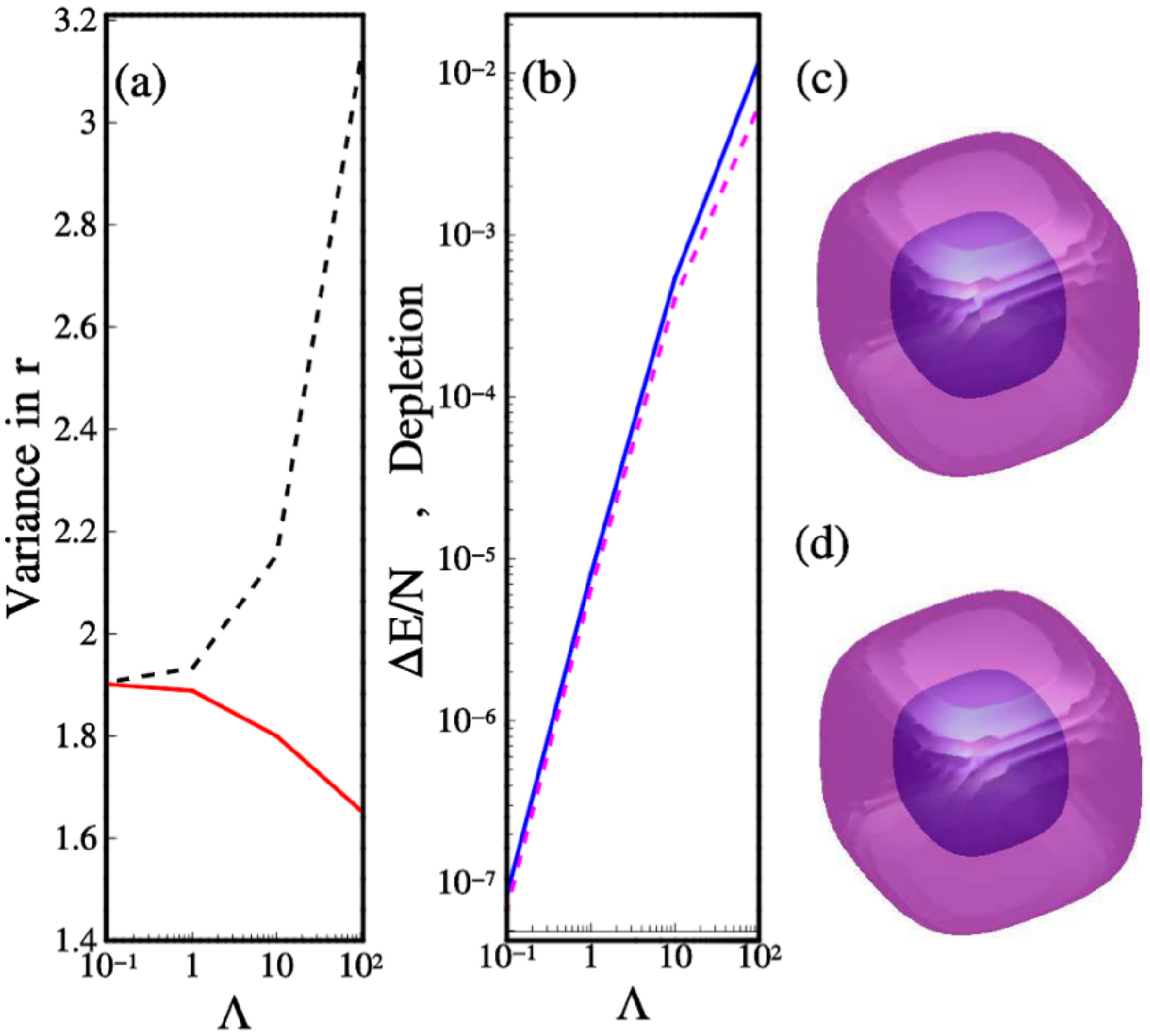}
\caption{(Color online) 
(a) The radial variance (in red; full curve) of a  three-dimensional Bose-Einstein condensate 
made of $N=100$ atoms in an cubic-shaped unharmonic trap as a function of the interaction parameter $\Lambda=\lambda_0(N-1)$.
The Gross-Pitaevskii variance is shown in black (dashed curve).
(b) The difference between the many-body and Gross-Pitaevskii energy (in blue; full curve)
and the depletion (in magenta; dashed curve).
(c) The many-body density and (d) the Gross-Pitaevskii density for $\Lambda=100$.
Note the depletion is less than $1\%$.
Shown are contours for $\frac{\rho(\r)}{N}=0.02$ and $0.001$.
The many-body density coincide with the Gross-Pitaevskii density,
whereas the respective variances deviate substantially from each other.
See text for further details.
The quantities shown are dimensionless. \\
}
\label{f5}
\end{figure}

The one-body Hamiltonian now takes on the from
$\hat h(\r) = -\frac{1}{2}\big(\frac{\partial^2}{\partial x^2}+\frac{\partial^2}{\partial y^2}+\frac{\partial^2}{\partial z^2}\big) + \frac{1}{36}(x^6+y^6+z^6)$.
The potential trap is isotropic and cubic-shaped.
The short-range repulsive interaction between the bosons is modeled by a
Gaussian function 
$\hat W(\r_1-\r_2)=\frac{1}{(2\pi\sigma^2)^{3/2}}e^{-\frac{(\r_1-\r_2)^2}{2\sigma^2}}$,
where $\sigma=0.25$
and the interaction parameters are 
$\Lambda=\lambda_0(N-1)=0.1$, $1$, $10$, and $100$.
The results are presented in Fig.~\ref{f5}.
As the interaction grows, the \textit{many-body} variance decreases whereas the Gross-Pitaevskii result increases,
see Fig.~\ref{f5}a.
Here we plot and compare the radial variances,
namely, $\frac{1}{N}(\Delta_{\hat X}^2+\Delta_{\hat Y}^2+\Delta_{\hat Z}^2)$
and $\Delta_{\hat x, GP}^2+\Delta_{\hat y, GP}^2+\Delta_{\hat z, GP}^2$.
Side by side, the depletion increases, see Fig.~\ref{f5}b.
Yet, even for the largest interaction we consider,
the depletion is $10^{-2}$,
 i.e., about $1$ atom is excited outside the first natural orbital.
Still, there is nearly a factor of one half
when comparing the variance
with the Gross-Pitaevskii result.
Interestingly,
the density, computed at the level of $M=4$ orbitals with MCTDHB,
see Fig.~\ref{f5}c,
and the Gross-Pitaevskii result,
see Fig.~\ref{f5}d,
hardly differ.
These densities clearly illustrate that, 
using the variance as a sensitive probe of many-body correlations,
there is much into BECs than meets the eye.
This concludes our investigations.

\section{Summary and Outlook}\label{Summary}

Many-body correlations are an important part of any trapped interacting Bose-Einstein condensate. 
Even in the limit of infinite particle number, where the energy and density have converged to their Gross-Pitaevskii counterparts, the effect of the correlations persists and can be directly observed in the variance of any operator. 
In this sense the variance of an operator serves as a highly sensitive probe of the many-body correlations in a BEC comprised of any particle number. 

Taking a BEC center-of-mass position as an example,
we demonstrated in a variety of systems that the variance of an operator
can substantially differ from that predicted by the Gross-Pitaevskii theory even when only one in a million particles are excited out of the condensate. 
All in all, the results and conclusions of this work are expected to open new perspectives for research of trapped BECs.

Further investigations on the dependence of the variance 
on the interaction strength and shape,
as well as its dependence on the topology of the trap, will shed new light on many-body correlations in BECs.
Studies of the variance of different observables,
such as the momentum and angular-momentum operators,
will provide further information on many-body
correlations of trapped BECs.
Last but not least,
it is inviting to extend the investigations
of the present work to the dynamics of BECs,
which could only offer more insight into the importance of many-body correlations \cite{KSMK}. 

\section*{Acknowledgements}

We thank Alexej Streltsov and Lorenz Cederbaum for discussions.
Partial financial support by
the DFG is acknowledged.
Computation time on the Cray XE6 system Hermit and
the Cray XC40 system Hornet 
at the HLRS are gratefully acknowledged.

\appendix
\section{Details of the Numerical Computations}\label{APP_VAR_NUMER}

The MCTDHB method is well documented 
in the literature.
It is based on the real-space Hamiltonian $\hat H(\r_1,\ldots,\r_N)$ and
utilizes $M$ time-dependent orbitals 
which are determined according to the variational principle.
As the method is variational,
propagation in imaginary time
convergences in the limit $M \to \infty$ orbitals 
to the ground state of the many-boson problem \cite{MCHB,Benchmarks}.
On the other end,
the Gross-Pitaevskii, mean-field theory is recovered for $M=1$.
In turn, utilizing $M$ time-dependent orbitals allows one to obtain accurate
numerical results with substantially less numerical resources
than a corresponding computation with $M$ fixed orbitals would need. 
The accuracy of the method has recently been demonstrated \cite{Benchmarks}. We use the implementation in \cite{package}.

\begin{figure}[!]
\includegraphics[width=0.8\columnwidth,angle=0]{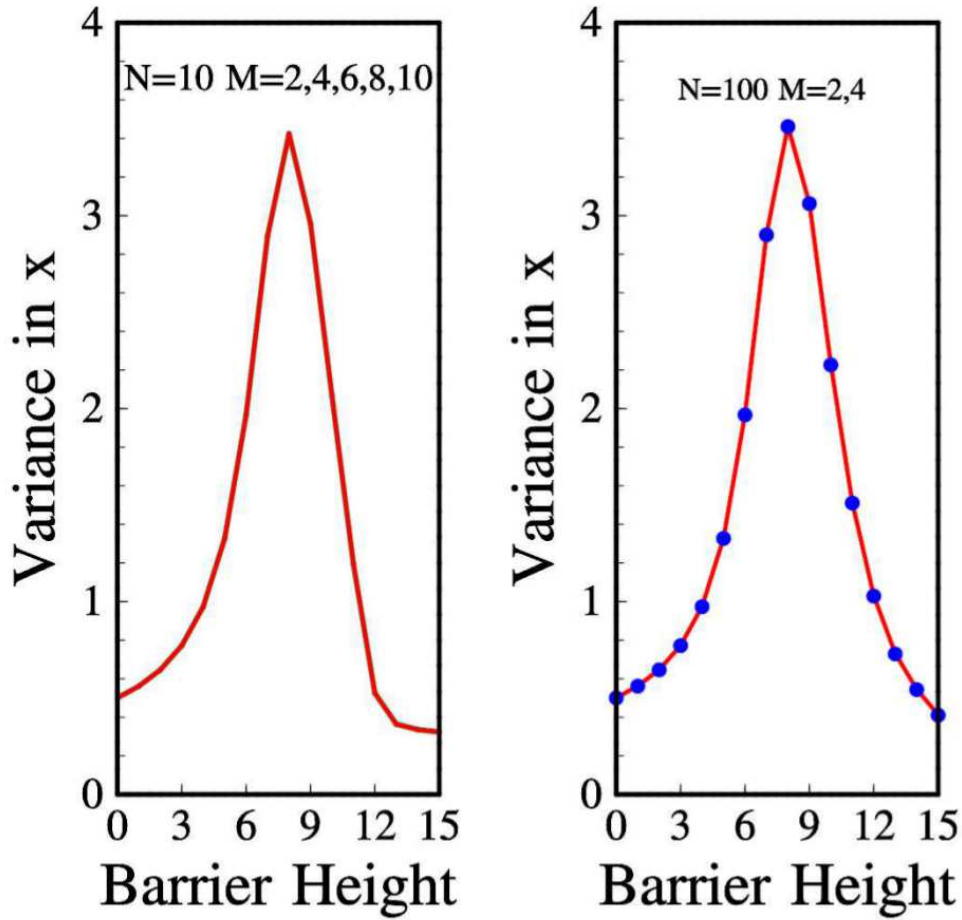}
\caption{(Color online) 
The variance of $\hat X$ of a Bose-Einstein condensate held in the symmetric trap
of Fig.~\ref{f1} for different barrier heights.
Results for $N=10$ and $100$ bosons for increasing
number $M$ of time-dependent orbitals are shown.
The curves sit atop each other,
and the variance seen to converge.
The quantities shown are dimensionless.}
\label{fA}
\end{figure}

In the present work the MCTDHB method is employed to compute the
ground state in a one-dimensional double well
(201 harmonic-oscillator discrete-variable-representation grid points
in a box of size [-20,20])
and in a three-dimensional cubic-shaped unharmonic trap
(64$^3$ grid points in a cube of size [-6,6]$\times$[-6,6]$\times$[-6,6] 
combined with a fast-Fourier-transform representation).
As a concrete example and without loss of generality,
convergence with increasing $M$ 
of the variance of $\hat X$ 
of a Bose-Einstein condensate held 
in the one-dimensional symmetric trap of Fig.~\ref{f1} 
is demonstrated in Fig.~\ref{fA}.

\end{document}